\documentclass{article}  
\usepackage{abstract_2e} 
\usepackage{amsmath}
\usepackage{times}
\usepackage{epsfig}

\newcommand{\jqsrt}{\textit{J. Quant. Spectrosc. Radiat. Transf.}}
\newcommand{\apjss}{\textit{Astrophys. J. Suppl. Ser.}}
\newcommand{\apj}{\textit{Astrophys. J.}}
\newcommand{\mnras}{\textit{Mon. Not. R. Astron. Soc.}}

\begin{document}  


\title{Completely Colorblind: Advances in Gray Techniques and Applications to Planets Near and Far}


\author{T. D. Robinson}
\affil{Department of Astronomy and Astrophysics, UC Santa Cruz, CA 95064, USA
(tydrobin@ucsc.edu)}


\runningtitle{Advances in Gray Techniques}

\titlemake  

\begin{abstracttext}
\section*{Introduction and History}
Gray approaches, which replace spectrally-resolved opacities with a 
wavelength independent mean opacity, have seen countless applications to 
problems of radiative transfer in stellar and planetary atmospheres since the 
original work of Schwarzschild (1906).  Applying the radiative equilibrium 
expressions of Schwarzschild (1906) to planetary thermal structure requires 
the assumption of an atmosphere that is essentially transparent to solar 
radiation, such that all shortwave absorption occurs at the surface or, for gas 
giants, at infinite depth.  These models, with different opacities for shortwave 
versus longwave radiation, are referred to as semi-gray, although the 
prefix is often dropped in the planetary literature.

Of course, planetary atmospheres are not transparent to shortwave 
radiation.  For Earth, water vapor is the main atmospheric absorber of 
solar radiation, and, in this context, Emden~(1913) and Milne (1922) were the 
first to explore how shortwave attenuation influences the radiative equilibrium 
thermal structure of an Earth-like atmosphere. Later, Hopf (1934) and 
Wildt (1966) explored the general solution to the semi-gray radiative equilibrium 
problem where both shortwave and longwave fluxes are attenuated.

McKay et al.~(1999) linearly combined the analytic model of Schwarzschild 
(1906) with those of Emden~(1913) and Milne (1922) to produce radiative 
equilibrium thermal structure profiles with both shortwave heating at the surface 
and in the upper atmosphere.  These models could develop stratospheric 
inversions, and were a very good match to the observed temperature structure 
of Titan's atmosphere, which is mostly in a state of radiative equilibrium.  
Building on these results, Robinson and Catling (2012) derived the first 
generalized analytic radiative-convective model for planetary atmospheres.  
This model is semi-gray, with two shortwave channels---one for heating the 
deep atmosphere and one for heating the upper atmosphere---and 
self-consistently determines where the radiative solution transitions to a 
convective adiabat.  Figure~\ref{fig:jupiter} shows the evolution of gray models, 
from Schwarzschild (1906) to Robinson and Catling (2012), as applied to Jupiter.

Today, the explosion of results and findings in the field of exoplanetary science 
has led to a resurgence of gray models for simulating planetary atmospheric 
structure.  Since exoplanet atmospheric characterization is, by and large, 
observationally limited, simple gray approaches can be much better constrained 
than complex, multi-parameter models.  As is discussed below, not only are gray  
techniques useful for providing constraints on atmospheric characteristics of 
exoplanets, but they also enable comparative studies of planets near and far by 
capturing the essential physics of planetary atmospheres in models that are 
broadly applicable and straightforward to interpret.

\begin{figure}
  \centering
  \vspace{5pt}
  \includegraphics[trim = 1mm 1mm 1mm 1mm, clip, width=7.5cm]{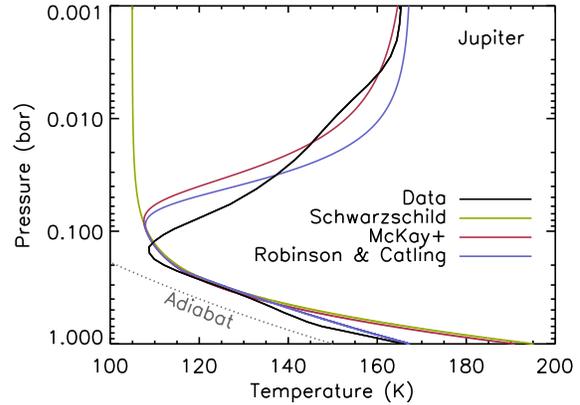}
  \caption{Applications of the Schwarzschild (1906), McKay et al.~(1999), and 
                Robinson and Catling (2012) models to Jupiter.}
  \label{fig:jupiter}
\end{figure}

\section*{Two Gray Approaches}
Gray techniques have been applied in both dynamic and equilibrium models.  For the 
former, pressure ($p$) dependent profiles of temperature ($T$) and 
atmospheric composition are used to determine the gray opacities of model levels 
at some timestep (using, e.g., a lookup table).  These opacities give the relationship 
between pressure and optical depth, and, thus, can be used when solving the 
two-stream equations of radiative transport (Schuster 1905; Schwarzschild 1906).  
Gradients in the net radiative flux then drive atmospheric heating and cooling, which 
are used to update the $T$-$p$ profile as the model timesteps forward.

Equilibrium models take a different approach.  Here, the two-stream equations are 
combined with an assumption of radiative (Schwarzschild 1906; Emden 1913; McKay 
et al.~1999) or radiative-convective (Robinson and Catling 2012) equilibrium to 
derive an analytic $\tau$-$T$ profile.  Given a $p$-$\tau$ relationship, either from an 
assumed parameterization or computed numerically, the equilibrium $T$-$p$ profile 
can be determined.  A commonly assumed relationship between optical depth and 
pressure is a power law, with $p \propto \tau^{n}$ (Pollack 1969).  As shall be 
discussed in the next section, the power $n$ has certain physical interpretations.

\section*{Gray Opacities}
Two key approaches to determining gray opacities exist.  First, gray opacities can 
be computed using spectrally-resolved gas (or grain) absorption spectra, commonly 
generated using line list databases (e.g., HITRAN; Rothman et al.~1987, 2013)
Such calculations are generally performed over a grid of 
pressure ($p$) and temperature ($T$) points, and assume some atmospheric 
chemical composition (from, e.g., thermal equilibrium chemistry).  The result is a 
lookup table of pressure and temperature dependent gray opacities.  It is uncommon 
to see gray opacities tabulated for terrestrial planetary atmospheres, as one cannot 
generally make the assumption that the atmosphere is a solar (or metal-enriched 
solar) composition gas in thermo-chemical equilibrium (which is typically used 
when computing gas giant opacities).  Thus, especially for terrestrial planets, it 
is common to heed the advice of Thomas and Stamnes (1999) who suggest 
determining gray opacities via comparisons between gray models and 
observations.  In this second key approach, one simply uses gray opacities as 
fitting parameters or as tools for comparative climatology.

When using spectrally-resolved models to compute gray opacities, a weighting is 
usually applied while integrating the resolved opacities ($\kappa_{\nu}$) over 
frequency.  One common averaging technique is the ``Planck mean'', where the 
mean opacity is determined using a weighting of the form $\kappa_{\nu}B_{\nu}(T)$, 
where $B_{\nu}$ is the Planck function.  This weighting emphasizes spectral regions 
near the peak of the Planck function where the resolved opacity is high, thereby 
ensuring that the gray flux \textit{emitted} by a thin atmospheric slab agrees with 
spectrally-resolved models.  A second averaging technique, called the ``Rosseland 
mean'', emphasizes regions near the peak of the Planck function where the resolved 
opacity is low, and uses a weighting of the form 
$\kappa_{\nu}^{-1} \mathrm{d}B_{\nu}(T)/\mathrm{d}\nu$.  This particular weighting 
ensures that the radiation diffusion limit is obeyed, thus making the Rosseland mean 
an appropriate choice deep in a planetary atmosphere (Mihalas 1970).

Recent tabulations of Planck and Rosseland mean opacities for brown dwarf and 
giant planet atmospheres can be found in Freedman et al.~(2008, 2014).
Figure~\ref{fig:mean_tau} shows examples of gray column optical depths through 
an isothermal Jupiter-like atmosphere for the Freedman et al.~opacities as well as a 
profile for Titan (from McKay et al.~1999).  Notice that the Freedman et al.~Rosseland 
mean optical depths, which are most sensitive to continuum collision-induced 
absorption (CIA) as well as Lorentzian line wings and apply best in the deep 
atmosphere, show a  $\tau \propto p^2$ scaling.  Both CIA and collisional line 
broadening are pressure dependent processes, such that one would expect 
$d\tau \propto p dp$ (Sagan 1969), as the gray opacity ($\kappa$) will be roughly 
proportional to pressure.  For the Freedman et al.~Planck mean opacities, which are 
most sensitive to Doppler broadened line cores and apply best in the upper 
atmosphere, we see a $\tau \propto p$ scaling.  This weaker dependence on pressure 
is due to the lack of pressure dependence in the Maxwellian distribution of molecular 
velocities, giving $\kappa$ roughly independent of $p$.  Note that $\tau \propto p$ 
and $\tau \propto p^2$ scalings are seen in the McKay et al.~Titan gray optical depth 
profile in the upper and lower atmosphere, respectively.

The analysis above would indicate that power law indexes for the $p \propto \tau^{n}$ 
relationship of roughly $n=$1--2 are most appropriate.  Indeed, McKay et al.~(1999) 
found that a value of $n=4/3$ produced a good fit of their gray thermal structure models 
to Titan's observed $T$-$p$ profile.  Note, however, that if the species primarily 
responsible for providing thermal opacity has a mixing ratio profile that varies with 
pressure, then larger or smaller values of $n$ might be expected.  For example, Weaver and 
Ramanathan (1995), when considering the small scale height for water vapor in 
Earth's atmosphere, suggest $n=4$.  However, Robinson and Catling (2014; their 
Figure S4) showed that using $n=2$ for models of gray thermal radiation in Earth's 
troposphere best reproduces thermal flux profiles from spectrally resolved models.

\begin{figure}
  \centering
  \includegraphics[trim = 2mm 1mm 2mm 3mm, clip, width=7.5cm]{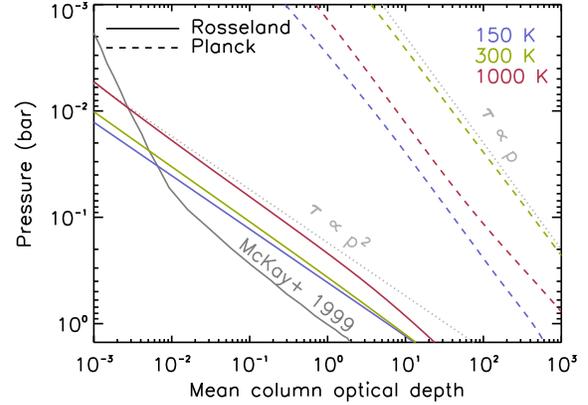}
  \caption{A selection of mean optical depths from the literature.  Power laws with 
                $\tau \propto p$ and $\tau \propto p^2$ are shown.  See text for more details.}
  \label{fig:mean_tau}
\end{figure}

\section*{Gray in 3-D}
The computational efficiency of gray radiative transfer has led to its adoption in 
numerous 3-D models of atmospheric circulation.  For example, in their studies of 
methane storms on Titan using 3-D models, Mitchell et al.~(2011) use a gray 
radiative transfer scheme with heritage in the McKay et al.~(1999) models.  Note 
that Mitchell et al.~assume a power law index of $n=3/2$, which is weighted more 
towards the deep atmosphere than the $n=4/3$ value used by McKay et 
al.~throughout the entire troposphere and stratosphere.

Frierson et al.~(2005) also adopt a gray radiation scheme in their simulations 
of moist processes in a 3-D Earth-like model.  Here, however, to  
more properly represent the dependence of opacity on pressure in both the 
troposphere and stratosphere, Frierson et al.~assume 
$\tau \sim f(p/p_{0}) + (1-f)(p/p_{0})^4$, where the parameter $f$ 
controls the transition from opacities being primarily dominated by water vapor 
and pressure broadening to being dominated by Doppler broadening, and $p_{0}$ 
is a reference pressure (at, e.g., the surface).  Of course, an added bonus from 
this parameterization is that radiative relaxation timescales in the stratosphere 
are much shorter, which can decrease model spin-up time.  More recently, 
Heng et al.~(2011) generalized this parameterization and applied it to 3-D 
studies of atmospheric circulation in Hot Jupiter atmospheres.

\section*{Gray Comparative Climatology}
The generality of gray techniques make them ideally suited to studies in 
comparative climatology and planetology.  For example, after developing 
an analytic semi-gray model of Titan's thermal structure and associated 
haze effects, McKay et al.~(1999) then used their anti-greenhouse models 
to study surface temperatures of a hazy early Earth.  More recently, 
Robinson and Catling~(2014) used an analytic semi-gray radiative-convective 
model to explain the common $\sim\!0.1$~bar temperature minimum in the 
thermal structure of Earth, Jupiter, Saturn, Titan, Uranus, and Neptune.

\begin{figure}
  \centering
  \includegraphics[trim = 1mm 1mm 1mm 1mm, clip, width=7.5cm]{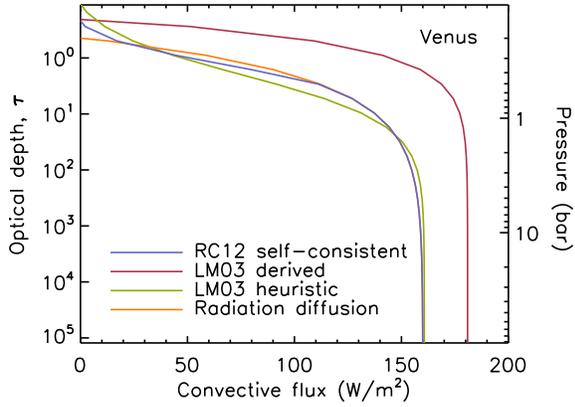}
  \caption{Convective flux in the Cytherean atmosphere from several gray 
                approaches. See text for more details.}
  \label{fig:venus}
\end{figure}

A common theme in comparative climatology works that use gray 
techniques is that of convection.   Sagan (1969) used gray and windowed-gray 
models to analyze the criteria for the onset of convective instability, which was 
revisited by Weaver and Ramanathan (1995) with an emphasis on the role 
that the vertical distribution of greenhouse gases plays in setting the steepness 
(and, thus, stability) of a radiative equilibrium temperature profile.  Lorenz and 
McKay (2003) used the Schwarzschild (1906) gray gas model to arrive at 
analytic expressions for the convective flux in a planetary atmosphere that is 
transparent to shortwave radiation.

\begin{figure}
  \centering
  \includegraphics[trim = 1mm 1mm 1mm 1mm, clip, width=7.5cm]{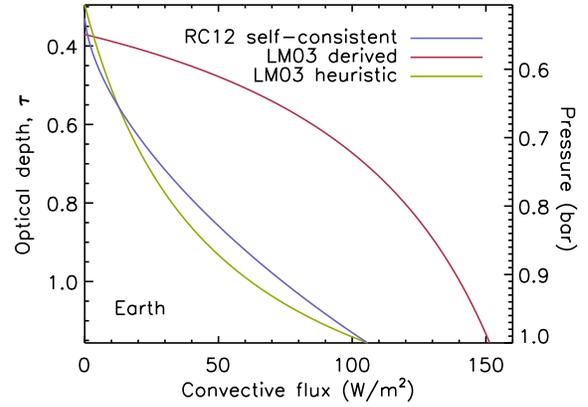}
  \caption{Convective flux in Earth's atmosphere from several gray 
                approaches. See text for more details.}
  \label{fig:earth}
\end{figure}

Figures~\ref{fig:venus} and \ref{fig:earth} show profiles of convective 
flux through the atmospheres of Venus and Earth, respectively, for several 
different models.  All models assume $n=2$ (which is appropriate for our 
focus on the deep atmospheres of these worlds), that all shortwave 
absorption occurs at the surface, and that the temperature profile in the 
convective portion of the atmosphere follows a dry adiabat.  The Robinson 
and Catling (2012; RC12) models are a self-consistent solution to the 
radiative-convective equilibrium problem, and adjoin a radiative equilibrium 
thermal structure profile to a convective profile by ensuring 
continuity of thermal radiative flux and temperature across the 
radiative-convective boundary.  The Lorenz and McKay (2003; LM03) 
``derived'' result has a convective flux profile that follows 
$F_{\rm{c}} = F_{\rm{s}}[ 1.5 - \frac{R_{\rm{s}}}{c_{p}}\frac{4}{nD\tau}(1+D\tau)]$, 
where $F_{\rm{s}}$ is the solar flux absorbed at the surface, $R_{\rm{s}}$ is 
the specific gas constant, $c_{p}$ is the atmospheric specific heat capacity, 
and $D$ is the so-called diffusivity factor that accounts for the integration 
of intensities over a hemisphere to arrive at the fluxes used in two-stream 
approaches.  We also show the Lorenz and McKay (2003) ``heuristic'' form 
of the convective flux, with 
$F_{\rm{c}} = F_{\rm{s}}(\tau - \tau_{rc})/[A + B(\tau - \tau_{rc})]$, where 
$A$ and $B$ are fitting parameters.  Note that we have adjusted this 
expression slightly to ensure that the convective flux goes to zero at 
the radiative-convective boundary (located at $\tau_{rc}$).  
Figure~\ref{fig:venus} also shows the analytic result for the radiation 
diffusion limit, where the net thermal flux is 
$\frac{2}{D}\frac{d\sigma T^{4}}{d\tau}$, so that (for an adiabatic 
temperature profile and with $\tau \propto p^{n}$) the convective flux is 
\begin{equation}
  F_{\rm{c}} = F_{\rm{s}} - \frac{8 R_{\rm{s}}}{n c_{p}} \frac{\sigma T_{0}^4}{D\tau_{0}}\left( \frac{\tau}{\tau_{0}} \right)^{\!\! \frac{4 R_{\rm{s}}}{n c_{p}}-1 } \ ,
\end{equation} 
where $T_{0}$ is the temperature at the surface (located at $\tau_{0}$).

The Venus and Earth cases both demonstrate that the derived result 
from Lorenz and McKay (2003) is not particularly good representations 
of the convective flux.  This likely stems from conflicting assumptions used 
when arriving at their expresson---the temperature gradient is taken as adiabatic 
while the temperature profile is assumed to be in radiative equilibrium.  The 
Lorenz and McKay (2003) heuristic model, which we fit to the self-consistent 
profile, yields much better convective flux profiles.  The functional form of this 
model is able to reproduce profiles ranging from the extremely optically thick 
Venus case to the relatively optically thin Earth case.

\section*{Conclusions and Future Work}
The simplicity of gray models makes them ideal for gaining intuition, while their 
generality makes these tools ideal for application to a broad range of planetary 
conditions.  It is no wonder that gray techniques and their application have been 
an active area of research for over 100~years.

A particularly exciting application of gray techniques that is currently seeing 
active development is in exoplanet retrieval analysis.  Here, gray radiative 
equilibrium thermal structure models are used to specify an atmospheric 
temperature profile with as few as four parameters (e.g., Line et al.~2012), 
and a radiative-convective equilibrium thermal structure could be specified with 
as few as 5--6 parameters.  This approach is especially attractive when 
attempting to minimize the number of parameters in a retrieval analysis, and 
offers an advantage over fitting for many tens of level dependent 
temperatures.

While gray models are currently seeing wide and diverse applications, a 
well known deficiency remains---gray radiative equilibrium 
profiles tend towards a constant ``skin temperature'' at low pressures.  This 
behavior is obviously unphysical, as was recently highlighted in a short paper 
on moist greenhouse atmospheres by Kasting et al.~(2015).  However, recent progress in 
developing gray-like models with ``picket fence'' opacities (Parmentier et 
al.~2015) have improved fits to atmospheric thermal structures at low pressures.  
Building on these results will enable the application of gray techniques to an 
even broader range of atmospheric conditions.

\section*{Acknowledgements}
TR gratefully acknowledges support from NASA through the Sagan Fellowship 
Program executed by the NASA Exoplanet Science Institute, and thanks R~Lorenz 
for thoughtful conversation on topics related to the history of gray techniques.

\end{abstracttext}

\end{document}